\documentclass[twocolumn, showpacs,superscriptaddress]{revtex4}
\usepackage{epsfig}
\usepackage{amssymb}
\usepackage{amsmath}
\usepackage{amsbsy}
\usepackage{cases}
\usepackage{color}
\usepackage{ulem}
\usepackage{epstopdf}

\begin{document}
\title{Miscibility Phase Diagram of Ring Polymer Blends: A Topological Effect}
\author{Takahiro Sakaue}
\email{sakaue@phys.kyushu-u.ac.jp}
\affiliation{Department of Physics, Kyushu University, Fukuoka 819-0395, Japan}
\author{Chihiro H. Nakajima}
\affiliation{WPI-AIMR, Tohoku University, 2-1-1 Katahira, Aoba-ku, Sendai, 980-8577, Japan}

\begin{abstract}
The miscibility of polymer blends, a classical problem in polymer science, may be altered, if one or both of the component do not have chain ends. Based on the idea of {\it topological volume}, we propose a mean-field theory to clarify how the topological constraints in ring polymers affect the phase behavior of the blends. While the large enhancement of the miscibility is expected for ring-linear polymer blends, the opposite trend toward demixing, albeit comparatively weak, is predicted for ring-ring polymer blends. Scaling formulas for the shift of critical point for both cases are derived. We discuss the valid range of the present theory, and the crossover to the linear polymer blends behaviors, which is expected for short chains. These analysis put forward a view that the topological constraints could be represented as an effective excluded-volume effects, in which the topological length plays a role of the screening factor.
\end{abstract}


\maketitle

\section{Introduction}
Ring polymers constitute a class of macromolecules, which do not possess chain ends~\cite{McLeish,Richter}. They could be constructed by just closing ends of linear polymers. Such a tiny difference in the molecular architecture, however, may influence the physical properties of polymers, which could be substantial in certain situations. Here, one speaks of the topological constraint, i.e., the molecules cannot spontaneously change their topology due to the non-crossability of bonds. 
Melts and concentrated solutions of ring polymers are
expected to be examples where the topological constraints manifest
themselves most drastically~\cite{RingMelt_Chromosome}.
%A Melt or concentrated solution of ring polymers would probably be an example, in which the most drastic effect of the topological constraint manifests~\cite{RingMelt_Chromosome}.

If we look at individual rings in the melt of non-concatenated (unlinked) and unknotted rings, their conformation is much different from that of Gaussian statistics expected for linear polymer counterparts.
Its clarification has been a subject of intense research for the last several decades~\cite{C-D,O-R-D,Muller1,Takano,Vettorel,Suzuki,SakauePRL,SakauePRE,Lang,Yoon,Grosberg_SoftMatter,Rosa_Everaers_PRL,Obukhov,Richter_PRL}.
Such a non-trivial conformation would affect the various macroscopic physical properties of the system.
The rheology of concentrated solutions would be one of primal examples in the list~\cite{Roovers,McKenna,Kapnistos,Milner,Halverson,Doi_Takano,Rubinstein_2016}. Another example is the phase behavior of polymer blends, which is the subject of the present study. 

The phase behavior, i.e., mixing or demixing in multi-component system, is a fundamental subject in various fields of condensed matter physics. It is also relevant to many of practical applications such as material engineerings and nutrition industry. Thus, seeking for a novel way to control the phase behavior is an interesting challenge.  This motivates us to clarify the topological effect, if any, in the phase behavior in ring polymer systems.
To get a feeling of the problem, it is instructive to consider the miscibility of the binary blend of ring polymers $A$ and linear polymers $B$, the former being unlinked and unknotted. The topological constraints in $A$-rings becomes tighter with the increase in their local concentration. As was pointed out by Khokhlov and Nechaev, this may lead to an enhanced compatibility in comparison with the corresponding linear polymer blends, since $B$-polymers play the role of a diluent softening the topological constraints in the homogeneous phase~\cite{Ring_linear_compatibility}. 
However, it is not obvious how to develop the above intuitive picture towards more quantitative predictions. One may also ask what happens in the ring-ring blends.

Our approach to the problem is to represent the constraints in ring polymers through the mean-field theory based on the idea of the {\it topological volume}~\cite{SakauePRL,SakauePRE}.  Such a strategy turned out to be rather successful to describe the topological constraints in single component system, which will be reviewed in Sec.~\ref{review} (with some revision). Although applying the same framework to blend involves some delicate point, we shall proceed with the simplest assumption. Most notably, we shall adopt the conventional Flory-Huggins theory to represent the non-topological part of the free energy. This allows us to investigate how the balance between the intermolecular unlinking and intramolecular unknotting constraints affects the phase behavior of ring-ring as well as ring-linear blends in a systematic way.
We shall identify several scenarios depending on the combination of molecular weights $N_A$, $N_B$ of respective components, the topological length (see below), and the type of the blend (ring-linear or ring-ring). 
Such an analysis puts forward a view that the topological constraints could be represented as an effective excluded-volume effects, in which the topological length plays a role of the screening factor.
In addition, our theory predicts that the topological constraints enhance the phase separation in the ring-ring blends; the opposite trend to the ring-linear blends.
 We discuss its physical origin, and derive scaling formulas on the shift of critical point in several different regimes. Finally, we critically reexamine the use of Flory-Huggins theory, and estimate the range of validity of the present theory.
%This allows us to investigate the phase behavior of ring-ring blends, in which the effect of the chain length can be analyzed in a systematic way. As we discuss below, the earlier treatment by Khokhlov and Nechaev is located as a special limit of our theory.
%More recently, a suggestion has been made that it can be a model system to unveil the large scale organization of interphase chromosome in cell nucleus.

{\it Topological Length}---.
As a rule, the topological effects manifest themselves in scale larger than some empirically known characteristic lengths. The simplest example would be the average size $R_0$ of zero-thickness (no excluded-volume) unknotted polymer ring (with $N$ segments of size $a$) in its dilute solution~\cite{GrosbergPRL};
\if0
\begin{subnumcases}
 {R_0 \simeq} 
 a N^{1/2}  & $(N < N_0)$ \label{R_0_ring_short}\\
 a N_0^{1/2}(N/N_0)^{\nu} & $(N \gg N_0)$
 \label{R_0_ring}
\end{subnumcases}
\fi
\begin{eqnarray}
 R_0 \simeq \left\{
\begin{array}{ll}
  a N^{1/2}  & (N < N_0) \\
 a N_0^{1/2}(N/N_0)^{\nu} & (N \gg N_0)
 \label{R_0_ring}
\end{array}
\right.
\end{eqnarray}
where $N_0 \sim 300$ and $\nu \simeq 0.59$ being the critical exponent for the self-avoiding walk.
Similarly, the average size of unknotted and unlinked polymer ring in their melt is $R \simeq a N^{1/2}$ for $N < N_e$ and $R \sim N^{1/3}$ for $N \gg N_e$, where $N_e \sim 100$ is an analog of the entanglement length familiar in the rheology of linear chain melts~\cite{Halverson}. In both cases, there is no real excluded-volume effect (either by the model definition or by the screening in the melt state). Nevertheless, the ideal (random walk) regime is restricted only to the small $N$ range, and the topologically controlled new regimes emerge for larger $N$. 
Note that the fact $N_e < N_c$ may be important when we discuss the onset of the topological effect in concentrated solutions. But, as will be shown below, in the evaluation of the free energy associated with topological constraints, the ratio $N_c/N_e$ just adds a constant factor, implying that the strict distinction between $N_0$ and $N_e$ seem not to be crucial in the current level of discussion. Whenever convenient, we will therefore absorb the ratio into the numerical coefficients in the scaling formulas.
%Since the strict distinction between $N_0$ and $N_e$ seem not to be crucial in our formulation below, we adopt the single notation $N_e $ for both of them, and call it the {\it topological length}, keeping in mind that the numerical values of $N_0$ and $N_e$ are not necessarily identical, and their relation remains unknown.

\section{Topological volume based mean-field theory}
\label{review}
\subsection{Basic formulation}
We first review the mean-field theory originally proposed to describe the size of individual molecules in concentrated solution of one-component rings~\cite{SakauePRL,SakauePRE}. Here, on one hand, rings tend to shrink their size to obey the unlinking constraint with their surrounding rings. But, on the other hand, the unknotting constraint within the ring itself acts against it. The equilibrium size of rings is determined by the competition between these two constraints. Such an approach dates back to a seminal paper by Cates and Deutsch~\cite{C-D}. Our topological volume based theory may be regarded as its revised version. For the clarity, we assume melt, i.e., the segment volume fraction $\phi \sim 1$ condition. $\phi$ dependence in semidilute solution will be discussed in Sec.~\ref{semi-dilute}.

The idea of the topological volume has been well documented in the context of the interaction between two closed, mutually unlinked rings in dilute solution. Here, the close approach results in the loss of conformational entropy, as some conformations are forbidden by the topology (unlinking constraint)~\cite{Frank-Kamenetskii}. Therefore, even in the absence of the segment repulsion, the second virial coefficient, i.e., an effective excluded volume takes a positive value on the order of the cube of the ring's size $R$. In concentrated solution, with the segment volume fraction $\phi$, we similarly assume the excluded-volume of topological origin scales as the cube of ring's size
\begin{eqnarray}
{\mathcal V} = R^3 {\mathcal Y}
\label{V_top}
\end{eqnarray}
where a factor ${\mathcal Y}$ (assumed to be independent of $N$ and $\phi$) accounts for the fact that, unlike a rigid impenetrable spheres, the polymer rings are very soft object, allowing a certain degree of  spatial overlapping:  rings are in an intense ``push and shove" ambiance.  One may then recall a sort of the packing problem, in which the free energy increases with the decrease in the ''free volume".
The simplest way to account for this would be, by following the idea of van der Waals, to assume the free energy per ring
\begin{eqnarray}
\frac{F_{unlink}(\Psi)}{k_BT}= - \ln{(1 - \Psi)}
\label{F_unlink}
\end{eqnarray}
which sharply diverges as the topological volume fraction $\Psi = {\mathcal V} {\mathcal X}/R^3 = {\mathcal X}{\mathcal Y}$ approaches the dense packed limit.
Here the coordination number ${\mathcal X} = R^3 /(N a^3)$ (also called as the overlapping parameter in the context of polymer entanglement) is the number of neighboring rings, which invades the volume $R^3$ of a reference ring.
%For further discussion, it may be useful to re-express the coordination number as ${\mathcal X} \simeq R^3/([N/g] \xi^3)$ with $\xi \simeq a g^{3/5} \simeq a \phi^{-3/4}$ and $g \simeq \phi^{-5/4}$, which indicates the view of semidilute solutions as a melt of concentration blobs. 

In this way, Eq.~(\ref{F_unlink}) represent a tendency that the rings favor shrinkage to satisfy the unlinking constraint. The more shrunk, however, the more severely the rings have to negotiate the unknotting constraint within individual rings. A simple dimensional analysis suggests the following form for this penalty
\begin{eqnarray}
\frac{F_{unknot}(\Psi; N)}{k_BT} \simeq 
\left(\frac{R_0}{R}\right)^{\beta}
\label{F_unknot_1}
\end{eqnarray}
where the reference size $R_0$ is given by Eq.~(\ref{R_0_ring}). The exponent $\beta$ can be determined by requiring the osmotic pressure to be intensive quantity, leading to $\beta = 6$ for $N<N_0$ or $3/(3\nu -1)$ for $N > N_0$. Here, we have re-derived the confinement free energy of the self-avoiding chain into cavity~\cite{Sakaue_Raphael, Grosberg_Khokhlov}, but with one difference, that is the presence of the topological length scale $N_0$~\footnote{The introduction of $N_0$ into the functional form of $F_{unknot}$ is the revised point in the present paper. In the original treatment in Refs.~\cite{SakauePRL} and~\cite{SakauePRE}, this point was missing, which amounts to say $\nu=1/2$.}.
Equation~(\ref{F_unknot_1}) is rewritten as
\begin{eqnarray}
&& \frac{F_{unknot}(\Psi; N)}{k_BT}  \nonumber \\
&&\simeq \left\{
%\begin{array}{ll}
%  N\left(\frac{ {\mathcal Y} }{ \Psi}\right)^{2}  & (N < N_0) \\
% N\left(\frac{ {\mathcal Y} }{ \Psi}\right)^{1/(3\nu-1)} N_0^{3(1-2\nu)/[2(3\nu-1)]}  & (N \gg N_0)
 \begin{array}{ll}
  N\left(\frac{ {\mathcal Y} }{ \Psi}\right)^{2}  & (N < N_0) \\
 N\left(\frac{ {\mathcal Y} }{ \Psi}\right)^{1/(3\nu-1)} N_0^{3(1-2\nu)/[2(3\nu-1)]}  &\\
\ =\left(\frac{\sqrt{N} {\mathcal Y} }{ \Psi}\right)^{1/(3\nu-1)} \left(\frac{N}{N_0}\right)^{3(2\nu-1)/[2(3\nu-1)]}  & (N \gg N_0)
 \label{F_unknot}
\end{array}
\right.
\end{eqnarray}
\if0
\begin{subnumcases}
 {\frac{F_{unknot}(\Psi; N)}{k_BT} \simeq} 
 N\left(\frac{ {\mathcal Y} }{ \Psi}\right)^{2}   & $(N < N_0)$ \label{F_unknot_short}\\
 N\left(\frac{ {\mathcal Y} }{ \Psi}\right)^{1/(3\nu-1)} N_0^{3(1-2\nu)/[2(3\nu-1)]} & $(N \gg N_0)$
 \label{F_unknot_long}
\end{subnumcases}
\fi
\if0
\begin{eqnarray}
\frac{F_{unknot}(\Psi; N)}{k_BT} \simeq   N\left(\frac{\phi {\mathcal Y} }{ \Psi}\right)^{1/(3\nu-1)} N_e^{3(1-2\nu)/[2(3\nu-1)]}
\label{F_unknot}
\end{eqnarray}
\fi
Each ring optimizes its size to reduce the total topological free energy $F_{top} =F_{unlink} + F_{unknot}$. 
%Here and in what follows, we omit the numerical constant in the scaling formula, which does not alter any qualitative conclusion.

We now determine the factor ${\mathcal Y}$ from the physical argument. 
When $\Psi$ is small enough $( \Psi < \Psi_e)$, the topological free energy becomes irrelevant in the thermal fluctuation. This observation is naturally connected to the topological length scale $N_e$.
From the condition $F_{unlink}(\Psi_e) \simeq F_{unknot}(\Psi_e; N_e) \simeq 0.5 \ k_BT$, one find $\Psi_e \simeq 0.4$ and can fix the softness factor
\begin{eqnarray}
{\mathcal Y} \simeq \frac{\Psi_e}{\sqrt{N_{e}}}
\label{Y-factor}
\end{eqnarray}
to be related to the entanglement length.
\if0
With Eq.~(\ref{Y-factor}), Eq.~(\ref{F_unknot_1}) is rewritten as
\begin{eqnarray}
\frac{F_{unknot}(\Psi; N)}{k_BT} \simeq   \frac{N}{N_e}\left(\frac{\Psi_e }{ \Psi}\right)^{1/(3\nu-1)}
\label{F_unknot_}
\end{eqnarray}
\fi
%, where in the last near-equality, we make use of the assumption commonly made in the linear polymer solution in good solvent that $N_e$ and $g$ have the same concentration dependence, and define the topological length $N_{e,melt}$ in the melt limit.

%{\it Mapping topological constraints to geometrical picture}---
For longer rings, the topological constraints becomes progressively important, which are to be optimized. 
One finds via the minimization of $F_{top}$ with respect to $\Psi (= \mathcal{X Y})$ that ${\mathcal X}$ slowly increases with $N$.
This yields $R \sim N^{1/3}$ for  $N > N_c$. But the crossover to this asymptotic is very slow $N_c \gg N_e$, and for medium length rings ($N_e \ll N \ll N_c$), the power-law fit may give the effective exponent close to $0.4$.

Equation~(\ref{Y-factor}) leads to 
%To describe such a topological effect in terms of the coordination number, let us rewrite Eq.~(\ref{F_unknot}) as $F_{unknot}/k_BT \simeq N/[g {\mathcal X}^2]$. Comparing this to the last expression in Eq.~(\ref{F_unknot}) with Eq.~(\ref{Y-factor}), we find
\begin{eqnarray}
{\mathcal X} = \sqrt{N_{e}} \   \Psi/\Psi_e
\label{X-Ne}
\end{eqnarray}
This is one of the key relations born of the present theory, which relate the geometrical quantity (${\mathcal X}$) and the topological length $N_{e}$, and bear some similarity to the so-called Kavassalis-Noolandi criterion on the entanglement length in dense linear polymer solutions (cf. Eq. (5) in Ref.~\cite{Kavassalis-Noolandi_PRL}). 
A rough estimate $N_{e} \sim 70$ (a typical value for flexible linear polymer melts) yields ${\mathcal X}_e \equiv {\mathcal X} (N_e) \sim 8$ at the onset of the topological effect toward an eventual saturation ${\mathcal X}_c \equiv {\mathcal X}(N_c) \sim 20$, which signals the onset of the compact statistics.
The onset length of the compact statistics is estimated as $N_c \simeq N_e {\mathcal X}_c^3/{\mathcal X}_e^2 \gtrsim 100 N_e$~\cite{SakauePRL, SakauePRE}.

At this point, it is instructive to gaze back at the outcome from the viewpoint of packing problem. Take randomly packed jammed particles, and ask what is the coordination number ${\mathcal X} $ (more precisely, the average number of contacts per particle). Insight from the marginal stability argument suggests that, for frictionless rigid particles, it  is equal to twice the number of degrees of freedom per particle, i.e., the isostatic conjecture.
%The following remark on the connection to general packing problem would be instructive at this point. The isostatic conjecture suggests that the average number of contacts per particle is equal to twice the number of degrees of freedom per particle for jammed randomly packed hard particles. 
This yields the average contact number $2 \times 6 = 12$ for an ellipsoid, which well approximates the average shape of individual rings in melts. The fact that the estimated value of ${\mathcal X} $ falls in the range suggested by isostatic argument may support the view of the concentrated ring polymer systems as a sort of the packing problem, where Eq.~(\ref{X-Ne}) provides a link to the topological constraint.

\subsection{Concentration dependence}
\label{semi-dilute}
In semidilute solution of rings, the topological lengths would become concentration dependent, which are denoted as $N_0^{(\phi)}$ and $N_e^{(\phi)}$. Here, we envision rings made from the succession of concentration blobs with size $\xi(\phi) \simeq a g(\phi)^{\nu} \simeq a \phi^{\nu/(1-3\nu)}$ (screening length of the excluded-volume effect). Then, $N_0$ would be replaced by $N_0^{(\phi)} \simeq N_0 g(\phi) \simeq N_0 \phi^{-1/(3\nu-1)}$. The ring size is now
\begin{eqnarray}
 R_0 \simeq \left\{
\begin{array}{ll}
   \xi(\phi) [N/g(\phi)]^{1/2} \simeq a \phi^{(2\nu-1)/[2(1-3\nu)]}  & (N < N_0^{(\phi)}) \\
 a N_0^{1/2}(N/N_0)^{\nu} & (N \gg N_0^{(\phi)})
 \label{R_0_ring_semi}
\end{array}
\right.
\end{eqnarray}
Keeping a factor $\phi$ in the coordination number ${\mathcal X} = R^3 \phi /(N a^3)$, and following the same line of argument from Eq.~(\ref{F_unknot_1}) to~(\ref{F_unknot}), we find
\begin{eqnarray}
&& \frac{F_{unknot}(\Psi; N)}{k_BT} \nonumber \\
&& \simeq \left\{
\begin{array}{ll}
  \frac{N}{g(\phi)}\left(\frac{ {\mathcal Y} }{ \Psi}\right)^{2}   & (N < N_0^{(\phi)}) \\
 \frac{N}{g(\phi)}\left(\frac{{\mathcal Y} }{ \Psi}\right)^{1/(3\nu-1)} N_0^{3(1-2\nu)/[2(3\nu-1)]}  & (N \gg N_0^{(\phi)})
 \label{F_unknot_semi}
\end{array}
\right.
\end{eqnarray}
By requiring $F_{unlink}(\Psi_e) \simeq F_{unknot}(\Psi_e; N_e^{(\phi)}) \simeq 0.5 k_BT$, we find Eqs.~(\ref{Y-factor}) and~(\ref{X-Ne}) for ${\mathcal Y}$ and ${\mathcal X}$, respectively, and 
\begin{eqnarray}
N_e^{(\phi)} \simeq N_e \phi^{-1/(3\nu-1)}
\label{N_e_phi}
\end{eqnarray}
With Eq.~(\ref{Y-factor}), one can rewrite $F_{unknot}$ as
\begin{eqnarray}
 \frac{F_{unknot}({\mathcal X}; N)}{k_BT} \simeq \left\{
\begin{array}{ll}
    \frac{N}{N_e^{(\phi)}}\left(\frac{ {\mathcal X}_e }{ {\mathcal X}}\right)^{2}   & (N < N_0^{(\phi)}) \\
 \frac{N}{N_e^{(\phi)}}\left(\frac{ {\mathcal X}_e }{ {\mathcal X}}\right)^{1/(3\nu-1)} & (N \gg N_0^{(\phi)})
 \label{F_unknot_semi_}
\end{array}
\right.
\end{eqnarray}
Note ${\mathcal X} ={\mathcal X}_e$ at $N = N_e^{(\phi)} \sim N_0^{(\phi)}$, which ensures a crossover between the above two expressions.

%The coordination number $X = \Psi/Y$ increases from $X(N_e)= Ne^{1/2}\phi^{5/8}$ at the onset of the topological effect $N=N_e$ and $\Psi={\tilde \Psi}$ toward the saturation $X(N^*) \rightarrow X(N_e)/{\tilde \Psi}$ in the compact statistics regime$N \rightarrow N^*$ and $\Psi \rightarrow 1$.

%$R \simeq a (N_e/{\tilde \Psi}^2)^{1/18} \phi^{-1/8}N^{4/9}$ and $R \simeq a (N_e/{\tilde \Psi}^2)^{1/6} \phi^{-1/8}N^{1/3}$ for moderate and very long rings, respectively.

\section{Blend with ring polymers}
\subsection{Free energy}
We aim at constructing the free energy of ring-ring blends, the suitable limit of which reduces to the free energy of the ring-linear blends.

{\it Ring-ring blend---.}
Consider a binary blend of A- and B-rings. The total numbers of respective rings are $M_A$ and $M_B$, which are contained in the volume $\Omega$. The chain length, the overall average composition of $\alpha$-ring are denoted as $N_{\alpha}$ and $\phi_{\alpha} =v_0 N_{\alpha}M_{\alpha}/\Omega$, where the monomer volume $v_0 \simeq a^3$ is assumed to be common to both types of ring, and $\phi_A + \phi_B =1$.

One complication in the blend is that $N_0$ and $N_e$, which are defined in the respective single component systems, can be different for $A$ and $B$ rings depending on their molecular structures. We assume here that these are common to $A$ and $B$ rings. With this simplifying assumption, we can construct the free energy associated with the topological constraints in the following way.

Let us first hypothetically switch off all the unlinking constraints in the blend. Then, the size $R_0$ of individual A- and B-rings is described by Eq.~(\ref{R_0_ring}), since the excluded-volume effect is screened at the monomer scale, and the unknotting constraint within individual rings induces swelling at $N>N_0$. In the real blend, however, the rings are squeezed by unlinking constraints, thus $R_{\alpha} < R_0$, where $R_{\alpha}$ is the size of $\alpha$-ring. 
Following the argument from Eq.~(\ref{F_unknot_1}) to~(\ref{F_unknot}), this squeezing leads to a free energy penalty per one $\alpha$-ring;
\begin{eqnarray}
\frac{F_{unknot}}{k_BT} &\simeq&
\left[N_{\alpha} \left(\frac{\phi_{\alpha} {\mathcal Y}_{\alpha} }{ \Psi_{\alpha}}\right)^{1/(3\nu-1)}\right]N_0^{3(1-2\nu)/[2(3\nu-1)]} 
\label{F_unknot_blend}
\end{eqnarray}
which is identified to be associated with the unknotting constraint. Here $\Psi_{\alpha}={\mathcal V}_{\alpha}M_{\alpha}/\Omega ={\mathcal X}_{\alpha} {\mathcal Y}_{\alpha}$ and ${\mathcal X}_{\alpha}= R_{\alpha}^3 \phi_{\alpha}/(N_{\alpha}a^3)$, and we focus on the case $N_{A},\ N_{B} > N_0 \sim N_e$ (otherwise, the topological effect would be negligibly small).

With the spatial size $R_{\alpha}$ of the $\alpha$-ring, we introduce the associated topological volume ${\mathcal V}_{\alpha}=R_{\alpha}^3 {\mathcal Y}_{\alpha}$.
The assumption of the common $N_{e}$ to both components implies the softness factor ${\mathcal Y}_A = {\mathcal Y}_B$ is common as well, which is given by Eq.~(\ref{Y-factor}).
The presence of the ``volume" ${\mathcal V}_{\alpha}$ implies the reduction of the associated ``free volume" $\Omega  \rightarrow \Omega - M_A {\mathcal V}_A - M_B {\mathcal V}_B$, yielding a factor $[1 - (M_A {\mathcal V}_A + M_B {\mathcal V}_B)/\Omega]^{M_A + M_B}$ in the partition function.

From above considerations, we obtain the free energies, associated with the unlinking and unknotting constraints, respectively, per monomeric volume as
\begin{eqnarray}
\frac{f_{unlink}}{k_BT}= - \frac{1}{N^*} \ln{(1 - \Psi)}
\label{f_unlink}
\end{eqnarray}
\begin{eqnarray}
\frac{f_{unknot}}{k_BT} &\simeq&
\left[\sum_{\alpha=A, B}\phi_{\alpha} \left(\frac{\phi_{\alpha} {\mathcal Y}_{\alpha} }{ \Psi_{\alpha}}\right)^{1/(3\nu-1)}\right]N_0^{3(1-2\nu)/[2(3\nu-1)]} \nonumber \\
&\simeq&\sum_{\alpha=A,B} \left[ \frac{\phi_{\alpha}}{N_e} \left( \frac{\phi_{\alpha} \Psi_e }{\Psi_{\alpha}}\right)^{1/(3\nu-1)}\right]
%\left[\phi_A \left(\frac{\phi_A {\mathcal Y} }{ \Psi_A}\right)^{1/(3\nu-1)} + \phi_B \left(\frac{\phi_B {\mathcal Y} }{ \Psi_B}\right)^{1/(3\nu-1)} \right]N_e^{3(1-2\nu)/[2(3\nu-1)]}
\label{f_unknot}
\end{eqnarray}
In the above equations, $N^{*}$ is a weighted harmonic mean of chain length defined as
\begin{eqnarray}
\frac{1}{N^{*}} \equiv \frac{\phi_A}{N_A} + \frac{\phi_B}{N_B}, 
\label{N_dagger}
\end{eqnarray}
and $\Psi = \Psi_A + \Psi_B$ is the total topological volume fraction. The use has been made of Eq.~(\ref{Y-factor}) to reach the last expression in Eq.~(\ref{f_unknot}).
%One caution deriving Eq.~(\ref{f_unknot}): unlike the one-component solution, the unperturbed ring size is $(R_{\alpha})_0 = a N_{\alpha}^{1/2}$ in the blend with overall melt condition $\phi_A + \phi_B = 1$. 
The total free energy is 
\begin{eqnarray}
f(\phi_A, \Psi_A, \Psi_B) = f_{FH} (\phi_A)+ f_{top}(\phi_A, \Psi_A, \Psi_B)
\label{f_total}
\end{eqnarray}
where $f_{top} = f_{unlink} + f_{unknot}$ and we adopt the conventional Flory-Huggins free energy
\begin{eqnarray}
\frac{f_{FH}}{k_BT} = \frac{\phi_A}{N_A}\ln{\phi_A}+\frac{\phi_B}{N_B}\ln{\phi_B} + \chi \phi_A \phi_B
\label{f_FH}
\end{eqnarray}
for the non-topological part with $\chi$ parameter to represent the nature of two-body interactions~\cite{deGennes}.

{\it Ring-linear blend---.}
For blends with ring (A) and linear (B) polymers, there is no contribution from linear polymers to the topological constraint, so we just need to set $\Psi_B = \phi_B =0$ in Eq.~(\ref{f_unlink}) and remove the unknotting contribution from B-component in Eq.~(\ref{f_unknot}); 
%\if0
\begin{eqnarray}
\frac{f_{unlink}}{k_BT}= -\left( \frac{\phi_A}{N_A} \right) \ln{(1 - \Psi_A)}
\label{f_unlink_RL}
\end{eqnarray}
\begin{eqnarray}
\frac{f_{unknot}}{k_BT} 
%= \phi_A \left(\frac{\phi_A {\mathcal Y}_A}{\Psi_A}\right)^{1/(3\nu-1)} N_e^{3(1-2\nu)/[2(3\nu-1)]}
\simeq \frac{\phi_A}{N_{e}} \left(\frac{\phi_A \Psi_e }{\Psi_A}\right)^{1/(3\nu-1)} \simeq \frac{\phi_A}{N_{e}^{(\phi_A)}} \left(\frac{ \Psi_e }{\Psi_A}\right)^{1/(3\nu-1)}
\label{f_unknot_RL}
\end{eqnarray}
%\fi
Note that $N_e (\sim N_0)$ in the above equation is the topological length in the melt of rings. In the presence of finite fraction of linear B-polymer, the onset of the topological effect on the conformation of A-ring should be delayed, hence, the corresponding length scale $N_e^{(\phi_A)}$ becomes longer. To determine $N_e^{(\phi_A)}$, we again require the condition $F_{unknot}(\Phi_e; N_e^{(\phi_A)}) \simeq 0.5 k_BT$ (see the argument around Eq.~(\ref{Y-factor}) and Sec.~\ref{semi-dilute}), where $F_{unknot} = f_{unknot} \times \Omega/(v_0 M_A)$ is the unknotting free energy per A-ring. One finds
\begin{eqnarray}
N_e^{(\phi_A)} \simeq N_e \ {\phi_A}^{1/(1-3\nu)}
\label{N_e_phi_A}
\end{eqnarray}
which leads to the last expression in Eq.~(\ref{f_unknot_RL}).

\subsection{Phase behavior of ring-linear blend}
\begin{figure}[h]
\includegraphics[width=0.34\textwidth]{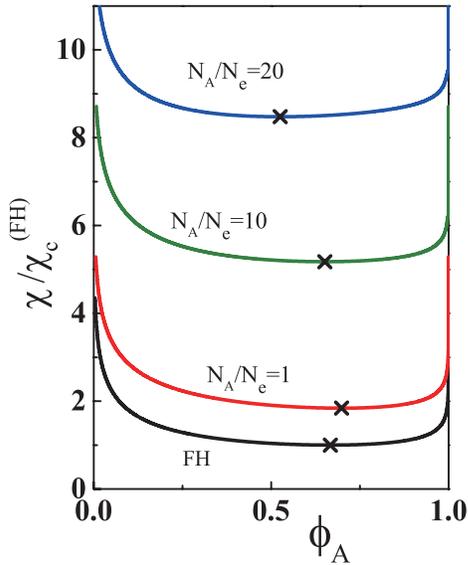}
\caption{Phase diagrams of the ring (A) - linear (B) polymer blends with the length ratio $\theta \equiv N_A/N_B = 1/4$ and $N_e=70$. Note that $\chi$ (vertical axis) is rescaled in unit of $\chi_c^{(FH)}$ so that the phase boundaries of linear polymer blends with a given $\theta$ but with different molecular weights superimpose a master curve (labeled as FH). The locations of critical point are marked by the cross symbols.
}
\label{Fig1}
\end{figure}

To find the equilibrium state, we first minimize $f_{top}$ with respect to $\Psi_A$. 
The optimum $\Psi_{A, min}$ is determined from the following algebraic equation
\begin{eqnarray}
g(\Psi_{A, min}) \simeq \frac{ N_A  }{  N_e}{\phi_A}^{1/(3\nu-1)}
\label{Psi_A_min}
\end{eqnarray}
with a function $g(x) \equiv x^{3\nu/(3\nu-1)}/(1-x)$.
It turns out that $f_{top}(\Psi_A=\Psi_{A, min})$ is a function of $\phi_A$ through the dependence of $\Psi_{A, min}$ on $\phi_A$. A graphical representation of Eq.~(\ref{Psi_A_min}) easily shows that the increase in $\phi_A$ leads to the increase in $\Psi_{A, min}$, which in turn results in the increase in $f_{top}$. This is the point, at which the topological constrains enter the problem of miscibility phase behavior. 
The phase diagrams calculated from our free energy is shown in Fig.~\ref{Fig1}.
Qualitatively, one can say that the system tends to avoid the high concentration state of ring polymers, which is unfavorable in terms of the topological constraints. The miscibility domain enlarges with the increase in the molecular weight.

%%%%%%%%%%%%%%%%%%%%%%%%%%%%%%%%%%%%%%%%%%%%%%%%%%%%%
%%%%%%%%%%%%%%%%%%%%%%%%%%%%%%%%%%%%%%%%%%%%%%%%%%%%%

We now develop the above qualitative picture to quantify the structure of the free energy.
Under the condition $N_A \gg N_e^{(\phi_A)}$, one finds $1-\Psi_{A,min} \simeq (N_e/N_A) \phi_A^{1/(1-3\nu)} \ll 1$, which approximates the topological free energy as
\begin{eqnarray}
\frac{f_{top}}{k_BT} \sim \frac{\phi_A}{N_A} \ln{\phi_A} + \frac{\phi_A^{3\nu/(3\nu-1)}}{N_e}
\label{f_top_RL_asymp}
\end{eqnarray}
aside from the term linear in $\phi_A$.
While the effect of the first term is just slightly to modify the translational entropy term in $f_{FH}$, the second term causes a qualitative change in the free energy profile $f(\phi_A)$. 

The critical point can be found from the usual procedure $\partial^2 f/\partial \phi_A^2 =\partial^3 f/\partial \phi_A^3=0$. One finds the critical composition $\phi_{A,c}$ and the critical $\chi_c$ as
\begin{eqnarray}
&&\phi_{A,c} \sim 
  \left( \frac{N_A}{N_e}\right)^{1-3\nu}    \label{phi_c_top}\\
&& \chi_c \sim \left\{
\begin{array}{ll}
  \frac{1}{N_A}\left( \frac{N_A}{N_e}\right)^{3\nu-1}  & \left( \frac{N_B}{N_e}  \gg \left(\frac{N_A}{N_e} \right)^{2-3\nu} \right) \\
\frac{1}{N_B} & \left( \frac{N_B}{N_e}  \ll \left(\frac{N_A}{N_e} \right)^{2-3\nu} \right)
\end{array}
\right.
  \label{chi_c_top}
\end{eqnarray}
\if0
One finds, under the condition $\theta \ll (N_A/N_e)^{2(3\nu-1)}$, the critical composition $\phi_{A,c}$ and the shift of the critical $\chi$ parameter $\Delta \chi_c \equiv \chi_c - \chi_c^{(FH)}$ as
\begin{eqnarray}
&&\phi_{A,c} \sim 
  \left( \frac{N_A}{N_e}\right)^{1-3\nu}    \label{phi_c_top}\\
&&\Delta \chi_c \sim 
  \frac{1}{N_A} \left(\frac{N_A}{N_e} \right)^{3\nu-1} + {\mathcal U}_{corr}
  \label{chi_c_top}
\end{eqnarray}
where the form of the correction term ${\mathcal U}_{corr}$ depends on the value of $\theta$. In the case of $\theta \gg 1$, the correction term, which takes the form  ${\mathcal U}_{corr} \sim - \theta^{1/2}/N_A$, becomes important. Indeed,
At $\theta  \rightarrow (N_A/N_e)^{2(3\nu-1)}$, the crossover to the normal scenario $\phi_{A,c} \sim \theta^{-1/2}$ and $\Delta \chi_c =0$ would be expected as elaborated in Sec. .
\fi

\if0
Physically, the condition II indicates the dominant contribution to the osmotic pressure $[\phi_A (\partial f/\partial \phi_A) -f]/v_0$ of the topological effect $\sim \phi_A^{3\nu/(3\nu-1)}$ over that of the third virial term $\sim \phi_A^3/N_B$ to locate the critical point. A similar consideration applies to the condition I as well, where the relevant quantity is the osmotic pressure $[\phi_B (\partial f/\partial \phi_B) -f]/v_0$ of the $B$-polymers. 
\fi

Compared to the above is the classical result from the analysis of the Flory-Huggins free energy~(\ref{f_FH}); 
\begin{eqnarray}
\phi_{A,c}^{(FH)}&=&(1+\theta^{1/2})^{-1} \label{phi_c_FH}\\
\chi_c^{(FH)} &=& [2 N_A (\phi_{A,c}^{(FH)})^2]^{-1} = \frac{(\sqrt{N_A} + \sqrt{N_B})^2}{2N_AN_B}
\label{chi_c_FH}
\end{eqnarray}
where $\theta = N_A/N_B$. 
For the critical point to be substantially shifted, the blend should be already topologically tight at $\phi_A = \phi_{A,c}^{(FH)}$. This condition $\Psi_{A,min}(\phi_{A,c}^{(FH)}) > \Psi_e$ is cast as
\begin{eqnarray}
N_A \left( 1+ \sqrt{\theta}\right)^{1/(1-3\nu)} \gg N_e
\label{condition_RL}
\end{eqnarray}

\subsection{Phase behavior of ring-ring blend}
%We assume $N_A \ge N_B$ without loss of generality.
As a new element in the ring-ring blend problem, one has to take account of the intense topological interaction between A-and B-rings in the blends, implying that the conformations of A-and B-rings are not independent in reducing the topological free energy.
Indeed, from $\partial f_{top}/\partial \Psi_A = \partial f_{top}/\partial \Psi_B =0$ (and $\phi_A + \phi_B=1$), we find
\begin{eqnarray}
\frac{\Psi_A}{\phi_A} = \frac{\Psi_B}{\phi_B} \ (= \Psi)
\label{Q_condition}
\end{eqnarray}
With this condition,  Eq~(\ref{f_unknot}) can be simplified as
\begin{eqnarray}
\frac{f_{unknot}}{k_BT} \simeq  \frac{1}{N_{e}}\left(\frac{\Psi_e}{\Psi}  \right)^{1/(3\nu-1)}
\label{f_unknot_2}
\end{eqnarray}

The optimum $\Psi_{min}$ is determined from the following algebraic equation
\begin{eqnarray}
g(\Psi_{min}) \simeq \frac{ N^*(\phi_A)  }{  N_e}
\label{Psi_min}
\end{eqnarray}
Unlike the ring-linear blend (cf. Eq.~(\ref{Psi_A_min})), the $\phi_A$-dependence of $\Psi_{min}$, thus $f_{top}$, is controlled by the $\phi_A$-dependence of $N^*$ in Eq.~(\ref{N_dagger}). For a symmetrical blend $N_A = N_B$, the $\phi_A$ dependence vanishes, resulting in no topological effect on the phase behavior.

The phase diagram calculated from our free energy is shown in Fig.~2. Contrary to the ring-linear blend, the miscibility is now {\it suppressed}, and the shift of phase boundary saturates at high molecular weight.

%%%%%%%%%%%%%%%%%%%%%%%%%%%%%%%%%%%%%%%%%%%%%%%%%%%%%
\begin{figure}[h]
\includegraphics[width=0.34\textwidth]{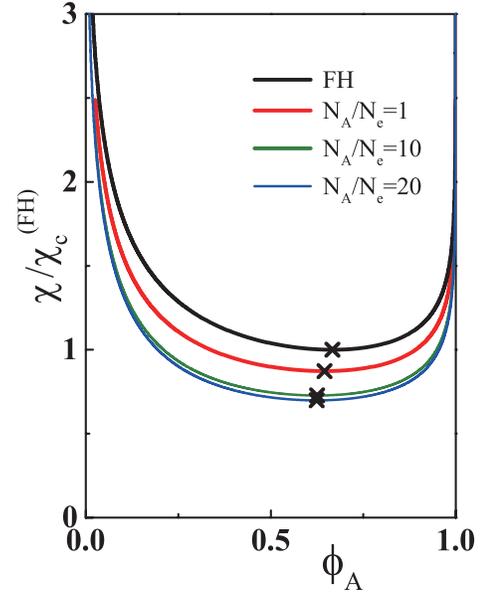}
\caption{Phase diagrams of the ring (A) - ring (B) polymer blends with the length ratio $\theta \equiv N_A/N_B = 1/4$ and $N_e=70$. The locations of critical point are marked by the cross symbols.
}
\label{Fig2}
\end{figure}
%%%%%%%%%%%%%%%%%%%%%%%%%%%%%%%%%%%%%%%%%%%%%%%%%%%%%

\if0
Inserting $\phi_{A,c}^{(FH)}$ into Eq.~(\ref{Psi_min}), we get the equation for $\Psi_{min}$ at the Flory-Huggins critical composition;
\begin{eqnarray}
g(\Psi_{min}|_{\phi_{A,c}^{(FH)}}) = \frac{c_1}{\theta -\sqrt{\theta}+ 1} \frac{N_A}{N_{e}}
\label{g_c}
\end{eqnarray}
For the critical point to be shifted, the topological free energy should exceed the thermal energy there, that is $\Psi_{min}|_{\phi_{A,c}^{(FH)}} > {\tilde \Psi}$, or in a more suggestive form 
\fi
To clarify the point, we analyze the structure of the free energy and the shift of the critical point.
Under the condition $1- \Psi_{min} \simeq N_e/N^* \ll 1$, the topological free energy is approximated as
\begin{eqnarray}
\frac{f_{top}}{k_BT} \sim \frac{1}{N^*}\ln{N^*}
\label{f_top_RR_asymp}
\end{eqnarray}
aside from the terms linear in $\phi_A$. 
The critical point is obtained as
\begin{eqnarray}
&&\phi_{A,c} \simeq \phi_{A,c}^{(FH)}  \\
&&\Delta \chi_c \simeq - \frac{\Delta^2}{2N_B(1- \phi_{A,c}^{(FH)} \Delta)} \sim \left\{
\begin{array}{ll}
  0  & (\theta = 1) \\
-\chi_c^{(FH)} & ( \theta  \gg 1, \ \theta \ll 1)
\end{array}
\right.
\label{critical_point_rr}
\end{eqnarray}
where $\Delta \chi_c \equiv \chi_c - \chi_c^{(FH)}$ and $\Delta \equiv (N_A-N_B)/N_A = 1- \theta^{-1}$.
The condition corresponding to Eq.~(\ref{condition_RL}) is
\begin{eqnarray}
\frac{N_A}{\theta -\sqrt{\theta} + 1} \gg N_{e}
\label{condition}
\end{eqnarray}

\if0
(I) (Nearly) symmetrical blends with long rings $(N_{e, melt} \ll N_B \simeq N_A)$: 
Although the condition~(\ref{condition}), which is now $N_A \simeq N_B >N_{e, melt}$, is well satisfied, the composition dependence of $N^*$ becomes very weak. This makes the topological effect on the phase diagram less prominent. 
In particular, as stated before, for the symmetrical blend $\theta = 1$, there is no shift in the phase behavior even the topological contribution to the free energy is substantially larger than the thermal energy. 
\fi

\section{Applicability of the Flory-Huggins approximation}
\label{Discussion}
So far, we have adapted the conventional Flory-Huggins theory for the non-topological part of the free energy without assessing its validity. In Flory-Huggins theory, polymers are supposed to overlap strongly in the concentration regime. In linear chain melt (in three dimensional space), chains take ideal Gaussian conformation and strongly overlapped with others~\cite{Daoud_Pincus}. In melt of rings, however, the Gaussian conformation is limited only up to the scale of $N_e$. In larger scales, the rings get more compact and exhibit a tendency of mutual segregation. This observation makes the use of Flory-Huggins theory questionable.

In this section, we will analyze the problem of the blend miscibility from (essentially the same but) a slightly different approach. In addition to re-deriving the results obtained so far, this analysis allows us to get the condition under which Flory-Huggins type approximation can be valid, and more importantly, to clarify various length scales relevant to the polymer collapse and phase-separation.

Let us assume $N_A \gg N_B$ and consider the A-ring polymers in the sea of B-linear polymers, and expand the free energy density (per monomeric volume $v_0$) as
\begin{eqnarray}
\frac{f}{k_BT} \simeq \frac{\phi_A}{N_A}\ln{\phi_A}+ \tau \phi_A^2 + B_3 \phi_A^3 + \frac{f_{top}}{k_BT}
\label{f_expand}
\end{eqnarray}
where (dimensionless) virial coefficients are $\tau=1/(2N_B)- \chi$, $B_3 = 1/(6N_B)$ from Eq.~(\ref{f_FH}), and we omitted an unimportant linear term in $\phi_A$. The factor $\sim N_B^{-1}$ in $\tau$ and $B_3$ represents a screening effect due to the B-polymer matrix~\cite{deGennes}. 
The osmotic pressure $\Pi$ is given by
\begin{eqnarray}
\frac{\Pi v_0}{k_BT} \simeq \frac{\phi_A}{N_A} + \tau \phi_A^2 + 2 B_3 \phi_A^3 + \frac{\Pi_{top} v_0}{k_BT}
\label{Pi_expand}
\end{eqnarray}
%with $\Pi_{top} v_0/k_BT \simeq \phi_A^{3\nu/(3\nu-1)}/N_e$.
 
In the following, we shall first review the phase behaviors of linear polymer blends. The purpose here is to identify the length scales (so called the thermal blob and the mesh size), and to see how these lengths are associated to the phase behaviors of the polymer blends. Then, we shall proceed to show how the topological effect alters those length scales. 
% As byproducts, we now have clearer view how the crossover to the conventional linear-blend type phase behavior takes place, and an indication that the topological constraints may be represented as an effective excluded-volume effect, where the topological length plays the role of screening factor.

\subsection{Reminder on linear-linear blend}
When two-body attraction becomes sufficiently large, the homogeneous state becomes unstable and the blend separates into A-rich and A-poor phases. The A-rich and-poor phases are stabilized by the three-body repulsive interaction and the translational entropy, respectively, which determines the respective phase boundaries. Therefore, by comparing the first and the third terms in osmotic pressure, we get the critical composition $\phi_{A,c} \simeq (N_B/N_A)^{1/2}$, which agrees with Eq.~(\ref{phi_c_FH}). To find the reduced temperature $\tau_c$ at the critical point, we look at a section of a ring with $n$ monomers, whose spatial extent is $r \simeq a n^{1/2}$. The free energy of that section is
\begin{eqnarray}
\frac{F_{n}}{k_BT} &\simeq& \tau v_0 \left( \frac{n}{r^3}\right)^2 r^3 + B_3 v_0^2  \left( \frac{n}{r^3}\right)^3 r^3 \nonumber \\
&\simeq& \tau  n^{1/2} + N_B^{-1}
\label{F_n_normal}
\end{eqnarray}
When $\tau<0$, the above equation leads to a length scale $\xi_{th} \simeq a g_{th}^{1/2} \simeq a/(|\tau| N_B)$ with 
\begin{eqnarray}
g_{th} \simeq (|\tau| N_B)^{-2}
\label{g_th}
\end{eqnarray} This so-called thermal blob signifies the scale, above which the effect of attractive interaction prevails. Another important length scale is the composition dependent mesh size, determined by the relations $a^3 g(\phi_A)/\xi^3(\phi_A) \simeq \phi_A$ and $\xi(\phi_A) \simeq a g^{1/2}(\phi_A)$, hence, $\xi(\phi_A) \simeq a \phi_A^{-1}$, $g(\phi_A) \simeq \phi_A^{-2}$. The homogeneous state is stable as long as $g_{th} > g(\phi_A)$. Thus, the condition $g_{th} = g(\phi_A)$ signifies the point at which the demixing takes place, thus, determines the phase boundary in A-rich side 
\begin{eqnarray}
|\tau| N_B \simeq \phi_A
\label{phase_boundary_normal}
\end{eqnarray}
At the critical concentration, this leads to $\tau_c \simeq -(N_A N_B)^{-1/2}$, which again agrees with the Flory-Huggins prediction Eq.~(\ref{chi_c_FH}).

\subsection{Topological effect in ring-linear blends}
For the topological contribution, we adopt the asymptotic form (Eq.~(\ref{f_top_RL_asymp}));
\begin{eqnarray}
\frac{f_{top}}{k_BT} \simeq \frac{\phi_A}{N_e^{(\phi_A)}} \simeq\frac{\phi_A^{3\nu/(3\nu-1)}}{N_e} \simeq \frac{\Pi_{top} v_0}{k_BT}
\label{f_top_expand}
\end{eqnarray}
From Eqs.~(\ref{f_expand}) and~(\ref{f_top_expand}), we see that the topological effect acts as the stabilization similar to the three-body repulsive term. If $\nu=1/2$, it has a cubic dependence on $\phi_A$, thus, acts as an effective third virial coefficient, and we would have a simple conclusion that the topological effect prevails when $N_e < N_B$. However, as suggested by Eq.~(\ref{R_0_ring}), the large scale behavior of a trivial knot without excluded-volume effect is described by the exponent of the self-avoiding walk $\nu \simeq 3/5$, and this makes the whole analysis more complicated as sketched below.

%%%%%%%%%%%%%%%%%%%%%%%%%%%%%%%%%%%%%%%%%%%%%%%%%%%%%
\begin{figure}[h]
\includegraphics[width=0.34\textwidth]{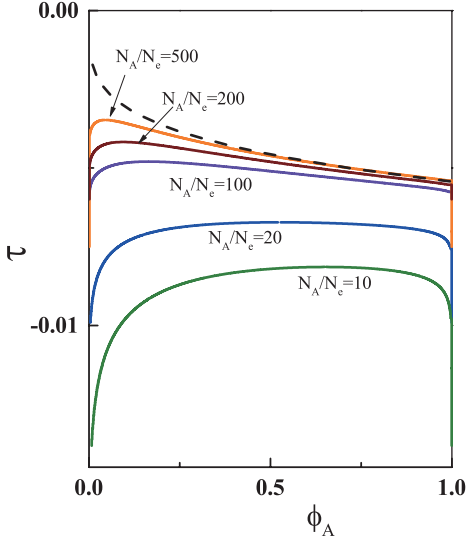}
\caption{Phase diagrams of the ring (A) - linear (B) polymer blends in $\phi_A - \tau$ plane. Parameters are the same as those in Fig.~\ref{Fig1}. Note that in this representation, the lower part (below co-existing curve) corresponds to the two-phase region. The dashed curve on the top is a guide to $|\tau| N_e \simeq \phi_A^{1/4}$, which is the asymptotic phase boundary (in high $\phi_A$ side) valid in long rings $N_A >N_e^{(\phi_A)}$.
}
\label{Fig3}
\end{figure}
%%%%%%%%%%%%%%%%%%%%%%%%%%%%%%%%%%%%%%%%%%%%%%%%%%%%%

We first look for the critical composition $\phi_{A,c}$. This can be done by comparing the first and the last terms in Eq.~(\ref{Pi_expand}), leading to the re-derivation of Eq.~(\ref{phi_c_top}).
We then look at the section (with $n$ monomers) of A-ring, and evaluate the corresponding free energy:
\begin{eqnarray}
\frac{F_{n}}{k_BT} 
%&\simeq& \tau v_0 \left( \frac{n}{r^3}\right)^2 r^3 + B_3 v_0^2  \left( \frac{n}{r^3}\right)^3 r^3 + \frac{1}{N_e v_0}\left( \frac{v_0 n}{r^3}\right)^{3\nu/(3\nu-1)}r^3\nonumber \\
\simeq \tau  n^{1/2} + N_B^{-1}+\frac{n^{q}}{N_e}
\label{F_n_top}
\end{eqnarray}
where the last term is evaluated from Eq.~(\ref{f_top_expand}) and we introduce an exponent $q=3(2\nu-1)/[2(3\nu-1)]$ just to simplify the notation (note that $q=3/8$ for $\nu=3/5$, while $q=0$ for $\nu=1/2$). Here it is assumed that the chain conformation is Gaussian at the length scale of interest, i.e., $n < N_0, N_e^{(\phi_A)}$.
Equation~(\ref{F_n_top}) suggests a length scale $g^* \simeq (N_e/N_B)^{1/q}$ beyond which the topological effect plays a dominant role for the stabilization. Under the situation where the demixing is controlled at the length scale larger than $g^*$, the thermal blob scale $\xi_{th}^{(top)} \simeq a [g_{th}^{(top)}]^{1/2}$ is determined by the balance between the first and the third terms in Eq.~(\ref{F_n_top}) with
\begin{eqnarray}
g_{th}^{(top)} \simeq (|\tau|N_e)^{2/(2q-1)}
\label{q_th_top}
\end{eqnarray}
As before, the phase boundary at A-rich side is obtained from the condition $g_{th}^{(top)} \simeq g(\phi_A)$;
\begin{eqnarray}
|\tau|N_e \simeq \phi_A^{1-2q}
\label{phase_boundary_top}
\end{eqnarray}
We find that, compared to the normal case~(\ref{phase_boundary_normal}), the topological effects lead to a peculiar shape of the phase boundary, which can be seen by rewriting the phase diagram (Fig.~\ref{Fig1} ) in $\phi_A - \tau$ plane, see Fig.~\ref{Fig3}. In addition, the role of $N_B$ as a screening factor is replaced by $N_e$.
By substituting the critical composition, we find the critical reduced temperature
\begin{eqnarray}
\tau_c N_e \simeq -  \left( \frac{N_e}{N_A}\right)^{2-3\nu}
\label{tau_c_top}
\end{eqnarray}
With the relation between $\tau$ and $\chi$ (see below Eq.~(\ref{f_expand})), we find its equivalence to Eq.~(\ref{chi_c_top}).

{\it Crossover to a classical scenario---.}
When the length of either ring or linear polymer is not long enough, the topological effect would not affect the phase behaviors of ring-linear blends substantially, and the crossover to a classical scenario is expected. The condition for the ring length is
\begin{eqnarray}
N_A > N_e^{(\phi_A)} \Leftrightarrow \phi_A > (N_e/N_A)^{3\nu-1}
\label{cross_over_FH_1}
\end{eqnarray}
where $N_e^{(\phi_A)}$ is given in Eq.~(\ref{N_e_phi_A}). This condition is required for the topological free energy to be approximated by its asymptotic form~(\ref{f_top_RL_asymp}).

The condition for the linear chain length is obtained as follows. 
By comparing $g_{th}$ and $g_{th}^{(top)}$, we find the condition $|\tau|N_e \simeq (N_B/N_e)^{(1-2q)/(2q)}$. This signifies the crossover to the Flory-Huggins scenario with negligible topological effects. Indeed, for the topologically controlled thermal blob to be meaningful, $g_{th}^{(top)}$ should be larger than $g^*$. This condition is satisfied only when 
\begin{eqnarray}
|\tau|N_e < (N_B/N_e)^{(1-2q)/(2q)}.
\label{cross_over_FH}
\end{eqnarray} 
Otherwise, the thermal blob follows a classical scaling (Eq.~(\ref{g_th})), which leads to the demixing described by the Flory-Huggins theory. On can check the above condition reduces $N_e < N_B$ if $\nu = 1/2$.

{\it Range of validity---.}
Our analysis above relies on the assumption of the Gaussian conformation up to, at least, the length scale $g(\phi_A), g_{th}^{(top)}$. Therefore, for the description to be valid, the condition $g(\phi_A), g_{th}^{(top)} < N_e^{(\phi_A)}, N_0$ should be required. We now bring the above condition into focus. 
First consider the case
\begin{eqnarray}
N_0 < N_e^{(\phi_A)} \Leftrightarrow \phi_A <  (N_e/N_0)^{3\nu-1}
\label{case_I}
\end{eqnarray}
In this case, the required conditions are
\begin{eqnarray}
&& \  g(\phi_A) < N_0 \Leftrightarrow N_0^{-1/2} < \phi_A \label{cond_a}\\
&& \  g_{th}^{(top)} < N_0 \Leftrightarrow |\tau|N_e > N_0^{(2q-1)/2} \label{cond_b}
\end{eqnarray}

In the case opposite to Eq.~(\ref{case_I}), i.e., $N_0 > N_e^{(\phi_A)} \Leftrightarrow \phi_A > (N_e/N_0)^{3\nu-1}$, thenA the condition $g(\phi_A) < N_e^{(\phi_A)}$ is automatically satisfied, so the required condition is 
\begin{eqnarray}
 && \ g_{th}^{(top)} < N_e^{(\phi_A)} \nonumber \\
  &&\Leftrightarrow |\tau|N_e > (N_e \phi_A^{1/(1-3\nu)})^{(2q-1)/2} \label{cond_c}
\end{eqnarray}
If (\ref{cond_a}) or (\ref{cond_b}) is not satisfied, a part of A-ring swells even inside the mesh or inside the thermal blob. On the other hand, if (\ref{cond_c}) is not satisfied, a part of A-ring starts to get compact even in the small scale where the attraction is yet a weak perturbation. In either case, one has to take account of the non-Gaussian conformation to discuss the blend phase behavior.

\begin{figure}[h]
\includegraphics[width=0.34\textwidth]{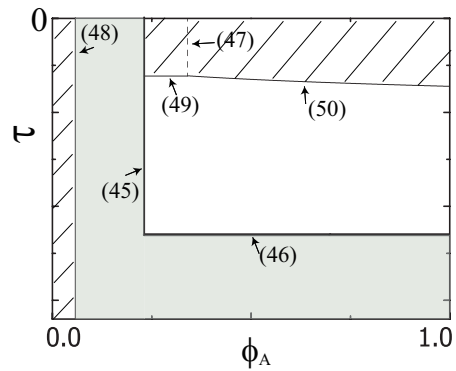}
\caption{Schematic diagram in $(\phi_A \ - \ \tau)$ plane illustrating regions for various regimes in ring-linear blend phase behaviors. Numbers in the figure refer to the corresponding equations in the text. The present theory would be applicable in the central white region, where the topological constraints are relevant and the non-topological part of the free energy can be taken into account through Flory-Huggins approximation. In the left and lower shaded region, the topological constraints are irrelevant, so the classical Flory-Huggins approximation would be valid for the description of blend phase behavior. In the top or the most left hatched region, the assumption of Gaussian conformation would not be appropriate to determine the thermal blob or the mesh size, where Flory-Huggins approximation needs to be modified to take into account the non-trivial statistics of binary contacts.
Using the typical numbers $N_e \sim 80$, $N_0 \sim 300$, $N_A = N_B =500$ and the exponent $\nu=3/5$, we have the following estimates for borders: (45) $\phi_A \simeq (N_e/N_A)^{4/5} \sim 0.23$; (46) $-\tau \simeq N_e^{-1} (N_B/N_e)^{1/3} \sim 0.023$; (47) $\phi_A \simeq (N_e/N_0)^{4/5} \sim 0.35 $; (48) $\phi_A \simeq N_0^{-1/2} \sim 0.05 $; (49) $-\tau \simeq N_e^{-1}N_0^{-1/8} \sim 0.006 $; (50) $-\tau \simeq N_e^{-9/8} \phi_A^{5/32} \rightarrow N_e^{-9/8} \ ( {\rm at} \phi_A \rightarrow 1) \sim 0.007$. For very long A-ring, the locations of the vertical lines (45) and (48) are interchanged. 
}
\label{Fig4}
\end{figure}

The above conditions in $(\phi_A \ - \ \tau)$ plane are summarized in Fig.~\ref{Fig4}.
The central white region exists when the vertical line Eq.~(\ref{cross_over_FH_1}) is located to the left-side of $\phi_A=1$ line and the border~(\ref{cond_b}) - (\ref{cond_c}) lies above the border~(\ref{cross_over_FH}). These condition can be expressed as $N_A > N_e$ and $N_B > N_e^{5/8}$. Note that the latter condition would be $N_B > N_e$ if the exponent introduced in Eq.~(\ref{R_0_ring}) was $\nu=1/2$.
%%%%%%%%%%%%%%%%%%%%%%%%%%%%%%%%%%%%%%%%%%%%%%%%%%%%%
%%%%%%%%%%%%%%%%%%%%%%%%%%%%%%%%%%%%%%%%%%%%%%%%%%%%%

\subsection{Topological effect in ring-ring blends}
For the topological contribution, we adopt the asymptotic form (Eq.~(\ref{f_top_RR_asymp})). 
\begin{eqnarray}
\frac{f_{top}}{k_BT} \sim  \frac{1-  \phi_A \Delta }{N_B} \ln{\left[\frac{N_B}{1- \phi_A \Delta}\right]}
\label{f_top_expand_RR}
\end{eqnarray}
where $\Delta \equiv 1-\theta^{-1}$ was defined below Eq.~(\ref{critical_point_rr}).
It is evident that there is no topological effect on the phase behavior for the symmetrical blends $N_A = N_B$.
In the case $\theta = N_A/N_B \gg 1$, we expand Eq.~(\ref{f_top_expand_RR}) into power series and obtain the total free energy density (see Eq.~(\ref{f_expand}))
\begin{eqnarray}
\frac{f}{k_BT} \simeq \frac{\phi_A}{N_A}\ln{\phi_A}+ {\tilde \tau} \phi_A^2 + {\tilde B}_3 \phi_A^3 
\label{f_expand_RR}
\end{eqnarray}
Therefore, we see that  the topological constraints affect shifting the effective virial coefficients 
\begin{eqnarray}
{\tilde \tau}&=&\tau-c_2\frac{\Delta^2}{N_B} \\
{\tilde B}_3 &=& B_3  - c_3\frac{\Delta^3}{N_B}.
\label{effective_virial}
\end{eqnarray}
Note that the unknown numerical constants $c_2, \ c_3$in the shift factors are due to the scaling estimate of the topological free energy~(\ref{f_top_expand_RR}).
The shift of $\tau$ in the negative direction indicates the enhancement of the phase separation. The equivalent expression in term of $\chi$-parameter is ${\tilde \chi} = \chi + c_2 \Delta^2/N_B$, which is consistent with Eq.~(\ref{critical_point_rr}).
Unlike the ring-linear blend case, the size of thermal blob follows the normal scaling, that is the number of monomers inside blob ${\tilde g}_{th} \simeq ({\tilde B}_3/|{\tilde \tau}|)^2$ (cf. Eq.
~(\ref{g_th})), and in this sense, the phase behavior of the ring-ring blends are qualitatively described by the normal Flory-Huggins theory with the shifted $\chi$-parameter.

{\it Range of validity---}.
In ring-ring blends, we may disregard the composition dependence of $N_e$ (at least, in the simplest situation we consider, where $N_e$ is common to both components). Since $N_e$ is numerically smaller than $N_0$,  the required conditions are: 
\begin{eqnarray}
\ &&g(\phi_A) < N_e \Leftrightarrow N_e^{-1/2} < \phi_A \label{cond_aa} \\
\ &&{\tilde g}_{th} < N_e \Leftrightarrow |{\tilde \tau}|N_e > {\tilde B}_3N_e^{1/2} \label{cond_bb} 
\end{eqnarray}
If (\ref{cond_aa}) or (\ref{cond_bb}) is not satisfied, one has to take account of the non-Gaussian conformation to discuss the blend phase behavior.

\section{Summary and perspectives}
In summary, the topological constraints in the blends with long ring polymers are generally relevant ingredients for their phase behaviors. The present theory, based on the idea of the topological volume, enables one to treat the ring-ring blends and the ring-linear blends on the same footing. In both cases, we have argued that it is the balance of unlinking and unknotting constraints that eventually affects the phase behaviors. This naturally allows us to quantify the disparity between ring-ring and ring-linear blends that leads to opposite trends for the shift of phase diagram. 

The general trend in ring-linear blends is the enhancement of miscibility. Here, the topological interactions among A-rings may play a dominant role to stabilize their dense phase. While this stabilization in normal case is realized by the three-body interactions whose strength is controlled by the inverse length $N_B^{-1}$ of B-polymers (Eqs.~(\ref{f_expand}) and~(\ref{F_n_normal})), the topological stabilization has a fractional power dependence on the composition and its strength is controlled by $N_e^{-1}$ (Eqs.~(\ref{f_top_expand}) and~(\ref{F_n_top})). This leads to a qualitative change in the shape of the A-rich side phase boundary, hence, a large shift of the critical point.
In ring-ring blend, on the other hand, the net effect of  the topological interactions is to shift the effective virial coefficients (hence, the $\chi$-parameter) in such a way that the miscibility will be suppressed. For $N_A \gg N_B$ case, this shift factor is controlled by $\Delta^2 N_B^{-1}$, implying that the relative shift of the critical $\chi_c$ is, at most, on the order of unity (see Eq.~(\ref{critical_point_rr})). The effect disappears for the symmetrical blend $\Delta=0$.
Note that the analytical predictions in Sec.~\ref{Discussion} are based on the asymptotic forms of topological free energy valid for sufficiently long chains (see Eqs.~(\ref{f_top_expand}) and~(\ref{f_top_expand_RR})). A prominent crossover behavior expected for medium length chains is clearly seen in Fig.~\ref{Fig3}.

It may be interesting to observe that the topological constraints could be represented as effective excluded-volume effects, where the topological length acts as the screening factor. Indeed, the size behavior in Eq.~(\ref{R_0_ring}) may be described by the following free energy
\begin{eqnarray}
\frac{F_{chain}}{k_BT} \simeq \frac{R^2}{Na^2} + {\tilde A}_2 \frac{N^2}{R^3}
\end{eqnarray}
with ${\tilde A}_2 = A_2 + a^3/\sqrt{N_0}$. Under the condition of negligible excluded volume $A_2 \simeq 0$, the topological term due to the unknotting constraint dominates the effective second virial coefficient. The screening factor $1/\sqrt{N_0}$ signifies the scale $N_0$ beyond which the perturbation due to the topological repulsion becomes substantial.
The topological volume due to the unlinking constraints introduced in Eq.~(\ref{V_top}) has the similar structure, i.e., the volume multiplied by a factor $\Psi_e/\sqrt{N_e}$ (which may be numerically close to $1/\sqrt{N_0}$).
As already summarized above, in the ring-linear blend, under the condition $N_A \gg N_e^{(\phi_A)}  \Leftrightarrow 1- \Psi \ll 1$, the balance between the unlinking and unknotting constraints leads to the free energy density, where $N_e$ plays an analogous role as the screening factor of excluded volume interactions.
\if0
In the earlier treatment of the ring-linear blend, only the term $f_{unknot}$ is considered as the topological constraint with an implicit assumption $\Psi_A = const$~\cite{Ring_linear_compatibility}. This amounts to assuming a relatively sharp transition of the size exponent $R \sim N^{1/2}$ to $R \sim N^{1/3}$ at $N=N_{e}$ in the one component unlinked and unknotted ring melt, which is known to be not the case at present. Our topological volume based theory is able to capture the prominent crossover behavior of $R$, for which the crucial is a gradual increase in $\Psi$, or equivalently the coordination number ${\mathcal X}$, with $N$. With this element equipped,  the present formalism should provide a more accurate description of the blend phase behaviors.
\fi

To focus on the most salient feature in the problem, we have considered the binary polymer blends with a common topological length $N_{e}$. In future work, it should be interesting to relax this condition, which may be important for various real situations. But the difference in $N_{e}$ may often be linked to the difference in the segmental properties (stiffness etc.), which implies much richer behaviors. Other interesting questions include the kinetics of the phase separation, which is expected to be different from the linear-linear blends, and the phase behavior in confined space, where the geometrical constraints matters, too. The rheology of the blends can be also controlled by the phase separation, the effect of which would be most prominent in the case of ring-linear blends. 
%Probably, numerical simulations should be an important tool to investigate these problems.
The experimental verification of the topological effect on the phase behaviors should be an interesting challenge~\footnote{Recently, Nagoya group has observed the topological effects on the phase behaviors of ring-linear and ring-ring blends, which are in qualitative agreement with our prediction (private communication with  A. Takano and Y. Doi). }. 
Indeed, experimental or numerical test of the prediction may serve as a touchstone for the relevance of the notion of topological volume in dense ring polymer solutions, which should be an important step toward our improved understanding of the topological constraints in polymer systems.

\acknowledgments 
T. S thanks A. Grosberg for discussion on the entropy associated with the unknotting constraint.
This work was supported by KAKENHI [Grant No.16H00804,``Fluctuation and Structure", Grant No.24340100, Grant-in-Aid for Scientific Research (B)], Ministry of Education, Culture, Sports, Science and Technology (MEXT), Japan and JSPS Core-to-Core Program (Nonequilibrium Dynamics of Soft Matter and Information).


\begin{thebibliography}{40}
\bibitem{McLeish}T. McLeish, Science, {\bf 297}, 2005 (2002).
\bibitem{Richter}D. Richter, S. Goo{\ss}en and A. Wischnewski, Soft Matter, Advance Article (2015).
\bibitem{RingMelt_Chromosome}J.D. Halverson, J. Smrek, K. Kremer and A. Y. Grosberg, Rep. Prog. Phys. {\bf 77}, 022601 (2014).
\bibitem{C-D}M.E. Cates and J.M. Deutsch, J. Physique {\bf 47}, 2121 (1986).
\bibitem{O-R-D}S. Obukhov, M. Rubinstein and T. Duke, Phys. Rev. Lett., {\bf 73}, 1263 (1994).
\bibitem{Muller1}M.M\"uller, J.P. Wittmer and M.E. Cates, Phys. Rev. E {\bf 53}, 5063 (1996).
\bibitem{Takano}A. Takano, Polym. Prepr. Jpn. {\bf 56}, 2424 (2007).
\bibitem{Vettorel}T. Vettorel, A.Yu. Grosberg and K. Kremer, Phys. Biol. {\bf 6}, 025013 (2009).
\bibitem{Suzuki}J. Suzuki, A. Takano, T. Deguchi and Y. Matsushita, J. Chem. Phys., {\bf 131}, 144902 (2009).
%\bibitem{Khokhlov_Nechaev}A.R. Khokhlov and S.K. Nechaev, Phys. Lett. A {\bf 112}, 156 (1985).
\bibitem{SakauePRL}T. Sakaue, Phys. Rev. Lett. {\bf 106}, 167802 (2011).
\bibitem{SakauePRE}T. Sakaue, Phys. Rev. E {\bf 85}, 021806 (2012).
\bibitem{Lang}M. Lang, J. Fischer and J.-U. Sommer, Macromolecules {\bf 45}, 7642 (2012).
%\bibitem{Lang_2}M. Lang, Macromolecules {\bf 46}, 1158 (2013).
\bibitem{Yoon}S.Y. Reigh and D.Y. Yoon, ACS Macro Lett. {\bf 2}, 296 (2013).
\bibitem{Grosberg_SoftMatter}A. Y. Grosberg, Soft Matter {\bf 10}, 560 (2014).
\bibitem{Rosa_Everaers_PRL}A. Rosa and R. Everaers, Phys. Rev. Lett., {\bf 112}, 118302 (2014).
\bibitem{Obukhov}S. Obukhov, A. Johner, J. baschnagel, H. Meyer and J.P. Wittmer, Europhys. Lett., {\bf 105}, 48005 (2014).
\bibitem{Richter_PRL}S. Goo\ss en, et al, Phys. Rev. Lett. {\bf 113}, 168302 (2014).


\bibitem{Roovers}J. Roovers, Macromolecules {\bf 18}, 1359 (1985).
\bibitem{McKenna}G.B. McKenna et al, Macromolecules {\bf 20}, 498 (1987).
\bibitem{Kapnistos}M. Kapnistos et al, Nature Mater. {\bf 7}, 997 (2008).
\bibitem{Milner}S.T. Milner and J.D. Newhall, Phys. Rev. Lett., {\bf 105}, 208302 (2010).
\bibitem{Halverson} J.D. Halverson, G.S. Grest, A.Y. Grosberg, K. Kremer, Phys. Rev. Lett., {\bf 108}, 038301 (2012).
\bibitem{Doi_Takano}Y. Doi et al, Macromolecules {\bf 48}, 3140 (2015).
\bibitem{Rubinstein_2016}T. Ge, S. Panyukov and M. Rubinstein, Macromolecules {\bf 49}, 708 (2016). 

\bibitem{Ring_linear_compatibility}A.R. Khokhlov and S.K. Nechaev, J. Phys. II France {\bf 6}, 1547 (1996).

\bibitem{GrosbergPRL}A.Yu. Grosberg, Phys. Rev. Lett. {\bf 85}, 3858 (2000).
%\bibitem{Deguchi}T. Deguchi and K. Tsurusaki, Phys. Rev. E {\bf 55}, 6245 (1997).
%\bibitem{Deguchi}T. Deguchi and K. Tsurusaki, Phys. Rev. E {\bf 55}, 6245 (1997).
\bibitem{Frank-Kamenetskii}M.D. Frank-Kamenetskii, A.V. Lukashin and A. V. Vologodskii, Nature {\bf 258}, 398 (1975).

\bibitem{deGennes}P.-G. de Gennes, {\it Scaling Concepts in Polymer Physics} (Cornell University Press, Ithaca, 1979).
%\bibitem{Doi_Edwards}M. Doi and S.F. Edwards, {\it The Theory of Polymer Dynamics} (Oxford University Press, Oxford, 1986).
\bibitem{Sakaue_Raphael}T. Sakaue and E. Rapha\"el, Macromolecules {\bf 39}, 2621 (2006).
\bibitem{Grosberg_Khokhlov} A. Grosberg and A. Khokhlov, {\it  Statistical Physics of Macromolecules} (AIP, NY, 1994).
\bibitem{Kavassalis-Noolandi_PRL}T.A. Kavassalis and J. Noolandi, Phys. Rev. Lett., {\bf 59}, 2674 (1987).
\bibitem{Daoud_Pincus}M. Daoud, P. Pincus, W.H. Stockmayer and T. Witten, Macromolecules {\bf 16}, 1833 (1983).
%\bibitem{Arrighi}V. Arrighi, et. al., Macromolecules {\bf 37}, 8057 (2004).

%\bibitem{Pakula}T. Pakula and S. Geyler, Macromolecules {\bf 21}, 1665 (1988).

%\bibitem{Brown}S. Brown, G. Szamel, J. Chem. Phys. {\bf 109}, 6184 (1998).
%\bibitem{Muller2}M.M\"uller, J.P. Wittmer and M.E. Cates, Phys. Rev. E {\bf 61}, 4078 (2000).
%\bibitem{Muller3}M.M\"uller, J.P. Wittmer and J.-L. Barrat, Europhys. Lett. {\bf 52}, 406 (2000).


%\bibitem{crumpled_globule}A.Yu. Grosberg, S.K. Nechaev and E.I. Shakhnovich, J. Phys. France {\bf 49}, 2095 (1988).
%\bibitem{crumpled_globule2}A.Yu. Grosberg, Y. Rabin, S. Havlin and A. Neer, Europhys. Lett. {\bf 23}, 373 (1993).

%\bibitem{chromosome_territories}E. Lieberman-Aiden, et. al., Science {\bf 326}, 289 (2009).
%\bibitem{chromosome_territories2}C. Lanct\^ot, et. al., Nat. Rev. Genet. {\bf 8}, 104 (2007).

%\bibitem{topological_swelling1}A. Dobay, J. Dubochet, K. Millett, P.-E. Sottas and A. Stasiak, Proc. Natl. Acad. Sci. USA {\bf 100}, 5611 (2003).

%\bibitem{Sakaue_Wada}T. Sakaue, G. Witz, G. Dietler and H. Wada, Europhys. Lett. in press.
%\bibitem{desCloizeaux}J.des Cloizeaux, J. Phys. Lett. {\bf 42}, L433 (1981).




%\bibitem{Deguchi}T. Deguchi and K. Tsurusaki, Phys. Rev. E {\bf 55}, 6245 (1997).


















\end{thebibliography}
\end{document}